\documentclass[aps,prl,twocolumn,floatfix,footinbib,showpacs,superscriptaddress]{revtex4-1}
\usepackage[english]{babel}
\usepackage[utf8]{inputenc}
\usepackage{graphicx}
\usepackage{graphicx}
\usepackage{amsfonts,amsmath,amssymb}
\usepackage{amsthm}
\usepackage{dsfont,bm}
\usepackage{color}
\usepackage{soul} 
\usepackage{amsbsy}
\usepackage[colorlinks=true,linkcolor=blue,pagecolor=blue,filecolor=blue,
menucolor=blue,urlcolor=blue, citecolor=blue,anchorcolor=blue]{hyperref}%

\begin{document}

\title{Spin-dependent thermoelectric effects in graphene based superconductor 
junctions}

\author{Razieh Beiranvand}
\affiliation{Department of Physics, K.N. Toosi University of Technology, 
Tehran 15875-4416, Iran}
\author{Hossein Hamzehpour}
\email{hamzehpour@kntu.ac.ir}
\affiliation{Department of Physics, K.N. Toosi University of Technology, 
Tehran 15875-4416, Iran}
\affiliation{School of Physics, Institute for Research in Fundamental Sciences (IPM), Tehran 19395-5531, Iran}


\begin{abstract}
Using the Bogoliubov de-Gennes formalism, we investigate the charge and spin-dependent
thermoelectric effects in superconductor graphene junctions. Results demonstrate that 
despite normal-superconductor junctions, there is a temperature dependent spin 
thermopower both in the graphene-based ferromagnetic-superconductor (F-S) and 
ferromagnetic-Rashba spin-orbit region-superconductor (F-RSO-S) junctions. 
It is also shown that 
in the presence of Rashba spin-orbit interaction, the charge and spin-dependent Seebeck
coefficients can reach to their maximum up to 3.5$k_B/e$ and 2.5$k_B/e$, 
respectively. Remarkably, these coefficients have a zero-point critical value 
with respect to magnetic exchange field and chemical potential. This effect 
disappears when the Rashba coupling is absent. These results suggest that 
graphene-based superconductors can be used in spin-caloritronics devices.

\end{abstract}
\maketitle

\section{Introduction}
After the discovery of graphene \cite{nov1}, a single layer of carbon atoms, 
the concept of Dirac fermions became more important for condensed matter 
physicists \cite{nov2,Demartino2007,Park2008}. The carriers of graphene in 
low-excitation regime are governed by 2D massless Dirac Hamiltonian and show a 
linear dispersion relation. The conduction and valance band in graphene touch 
each other in the Dirac points. Because of this unique geometrical structure, a 
wide variety of applications in electronics \cite{nov2,nov3}, opto-electronics 
\cite{Bonaccorso2010,Avouris2010,Bao2012} and even spintronics 
\cite{Zareyan2010,Pesin2012,Han2014,Ghosh2011} were implemented for graphene in the last 
decade. Transport properties of massless Dirac fermions in graphene display many 
 interesting behaviors such as anomalous quantum Hall effect 
\cite{Zhang2005,Novoselov2007,Gusynin2005,Rickhaus2012}, chiral tunneling 
\cite{Katselson2006,Beenakker2008}, Klein paradox 
\cite{Katselson2006,Zareyan2008} and so on \cite{Malec2011}. Furthermore, graphene acquires both 
ferromagnetic and superconducting features by means of proximity 
\cite{Buzdin2005,Heersche2,Wang2015}. These effects open a new opportunity for 
designing devices which are based on hybrid structures of superconductors. Since 
the graphene-based junctions are assumed as mesoscopic systems, one can 
employed the Landauer formalism to describe the transport properties of them in 
ballistic regime. Extending this formalism leads to Blonder-Thinkham-Klapwijk 
(BTK) approach \cite{BTK1982} for describing quantum transports of the system. 

The thermoelectric power or Seebeck coefficient and thermal current of graphene 
based junctions have been topics of intense research in few years 
\cite{Rameshti2015,Ghosh2015,Inglot2015,Dragoman2007,Xu2010,Salehi2010,Salehi20102}. 
Because of some experimental limitations in nano-scale devices on 2D materials, 
the charge and spin-dependent Seebeck effects are often ignored in the previous studies 
\cite{Uchida2008,Jaworski2012}. Recently, developing in the low-temperature 
measurement devices provide suitable condition for experimental observation in 
the field of thermal transport. More recently, theoretical and experimental 
investigations were done on the thermoelectric features of graphene by Zuev et 
al. \cite{Zuev2009} and Wei et al. \cite{Wei2009} demonstrating that the sign of 
Seebeck coefficient is changed due to the change in carrier type from electron 
to hole. In fact, we need electron-hole asymmetry at Fermi level to reach higher 
efficiency in thermoelectric power \cite{Ozaeta2014}. In graphene-based devices, 
chemical potential can be tuned close to the band edges which provide conditions 
to create asymmetry in density of states (DOSs). Applying spin-splitting field 
$h$ can also break the symmetry for each spin directions. In spin-splitting 
systems, the spin-polarized current can be generated by applying a temperature 
gradient. Usually, the combination of spin-orbit interactions (SOI) and a 
uniform Zeeman field create a useful mechanism to manipulate the quasi-particles 
transport which lead to very interesting phenomena in the field of spintronics 
\cite{Rashba2009,Kane2005,Beiranvand2016}. It is well known that SOI can be 
divided into two categories named Rashba spin-orbit interaction (RSOI) which is 
due to the structure inversion symmetry and Dresselhaus spin-orbit interaction 
(DSOI) which is the results of the bulk inversion symmetry \cite{Kane2005}. It 
was experimentally demonstrated that a graphene nano-sheet can support strong 
RSOI about 17 meV by proximity \cite{Avsar2014} while the strength of the DSOI 
is very small approximately between  0.0011 and 0.05 meV 
\cite{Min2006,Huertas2006,Yao2007}. To demonstrate the existence of 
proximity-induced RSOI in the graphene layer, a DC voltage along the graphene 
layer is measured by spin to charge current conversion which is interpreted as 
the inverse Rashba-Edelstein effect. This property can be employed as a means to 
control the spin-transport in the graphene-based spintronic devices.

In addition to the above, spin-caloritronics which is a combination of 
thermoelectric and spintronic effects has attracted more attention due to 
very promising applications \cite{Rameshti2015,Ghosh2015,Inglot2015,Lopez2014,Alomar2015}. 
In a non-magnetic material like graphene, generating the spin-dependent Seebeck effect 
which can produce an spin current with ability to convert into a measurable 
voltage is a very important issue. This spin -dependent Seebeck effect has been previously 
reported for metallic, semiconductor and even insulator materials 
\cite{Uchida2008,Uchida2010_1,Uchida2010_2,Qu2013}. The heating power applied to 
the system can easily change the accumulation of majority and minority spins and 
switch the sign of spin-dependent Seebeck coefficient with reversing the temperature 
differential. The results revealed a large number of possible applications in 
designing thermo-spintronic devices and increasing their efficiency 
\cite{Hubler2012,Hwang2016,Kolenda2016,Silaev2015}. Considering this fact that 
graphene has tunable electronic properties with strong energy-dependence of the 
conductivity along with very weak spin relaxation made it promising candidate in 
spin-thermoelectrics in comparison with other magnetic materials.

Herein, we consider two different setups of graphene-based junctions which have 
a potential for developing superconducting devices with interesting 
thermoelectric abilities under various condition of temperature, magnetic 
exchange field, chemical potential and spin-orbit interaction. We theoretically 
investigate the thermoelectric properties of ferromagnetic-superconductor (F-S) 
and ferromagnetic-Rashba spin-orbit region-superconductor (F-RSO-S) junctions 
made of graphene. Results demonstrate that the simultaneous effect of 
spin-splitting field and spin-orbit coupling lead to a strong suppression of the 
spin imbalances which can affect on the usually small thermoelectric 
coefficients in superconducting heterostructures. Also, the calculated spin 
Seebeck coefficient can easily tunned both through applied magnetic exchange 
field and chemical potential. Making accurate measurement of charge and spin 
Seebeck coefficients can lead to suggestion a new setup of graphene-based 
nano-structures in cooling apparatuses, thermal sensors or even renewable energy 
applications.

The rest of this paper is organized as follows. In Sec. II, we first express 
the theoretical formalism of our calculations. The results are discussed in Sec. 
III in two subsections: in Subsec. IIIA, we study the thermoelectric properties 
of F-S junction. Then the junction consist of RSO coupling, F-RSO-S, is 
discussed in Subsec. IIIB. We finally summarize concluding remarks in Sec. IV.

\section{THEORETICAL FORMALISM}\label{secII}
In order to describe the thermoelectric effects in our supposed systems, we 
implement the generalized BTK approach \cite{BTK1982}. By keeping the two ends 
of the junction at different temperatures, the charge and thermal currents flow 
across the junction:
\begin{equation}
\left( \begin{array}{cc}
I_{ch}\\
I_{th} \\
\end{array} \right) =\dfrac{W}{\pi^2\hbar}\int d\varepsilon \left( \begin{array}{cc}
\mathcal{G}\\
\mathcal{K} \\
\end{array} \right) N_\sigma(\varepsilon)[f_L(\varepsilon)-f_R(\varepsilon)] \;.
\label{Eq.current}
\end{equation}
These currents depend on the electric 
($\mathcal{G}=\mathcal{G}_\sigma+\mathcal{G}_{\bar{\sigma}}$) and thermal 
($\mathcal{K}=\mathcal{K}_\sigma+\mathcal{K}_{\bar{\sigma}}$) conductances in 
which $\sigma$ denotes the spin degree of freedom due to the presence of 
ferromagnetic features ($\sigma=\pm$ or $\uparrow\downarrow$ and 
$\bar{\sigma}=-\sigma$). In the standard BTK approach, the charge and 
thermal conductances for each spin direction obtained through the following 
equations \cite{Yokoyama2008,BTK1982}:

\begin{equation}
\begin{array}{cc}
\mathcal{G}_{\sigma}=g_0\int_{-\pi/2}^{\pi/2} \cos\alpha d\alpha   
\Big(1-\sum_{\sigma'=\pm} 
(|r_{N}^{\sigma'}|^2-|r_{A}^{\sigma'}|^2 )\Big) \;,
\\
\mathcal{K}_{\sigma}=g_0 \int_{-\pi/2}^{\pi/2}\cos\alpha d\alpha 
\Big(1-\sum_{\sigma'=\pm} 
(|r_{N}^{\sigma'}|^2+|r_{A}^{\sigma'}|^2) \Big) \;,
\end{array}
\label{G}
\end{equation}
where the spin-dependent normal and Andreev reflections ($r_N^{\sigma}, 
r_A^\sigma$) are obtained by matching the wave functions at the boundaries  
\cite{Beiranvand2016} and $g_0={2e^2}/{\hbar}$ represents the spin-dependent 
ballistic conductance of the junction as a function of particle's energy, 
$\varepsilon$. The term $f(\varepsilon)$ is the Fermi-Dirac distribution 
function, $W$ denotes the width of the junction and $N_\sigma(\varepsilon)$ is 
the spin-dependent density of state (DOS). 

In a system under temperature difference $\delta T$, the Seebeck coefficient 
is defined by
\begin{equation}
S_{\sigma}=-\Big(\dfrac{ V}{\delta T}\Big)_{I_{ch}=0} \;,
\end{equation}
where $V$ is the voltage derived at zero charge current in response to the 
temperature gradient.

In the linear response regime, the bias voltage $V$ and temperature difference 
$\delta T$ are assumed to be very small. Then, the expansion of the Fermi-Dirac 
distribution function can be written up to linear terms as,
\begin{equation}
\begin{array}{l}
f_{T-\delta T}(\varepsilon-eV)\approx f_T(\varepsilon)-eV\dfrac{\partial 
f}{\partial \varepsilon}+\dfrac{\partial T}{T}\varepsilon\dfrac{\partial 
f}{\partial \varepsilon} \;, 
\\ \\
f_{T+\delta T}(\varepsilon)\approx f_T(\varepsilon)-\dfrac{\partial 
T}{T}\varepsilon\dfrac{\partial f}{\partial \varepsilon}	 \;,
\\
\\
\dfrac{\partial f}{\partial\varepsilon}=\dfrac{-1}{4k_BT\cosh^2(\dfrac{\varepsilon}{2k_BT})}.
\end{array}
\end{equation}
So, the electric current in Eq. \ref{Eq.current} can be decoupled in two terms 
as
\begin{equation}
{I}_{ch}\approx GV+I_{T}\dfrac{\delta T}{T} \;.
\end{equation}

From the above equations, the linear thermopower for each spin direction is then 
given by
\begin{equation}
S_{\sigma}=-\frac{I_T^\sigma}{(G_\sigma+G_{\bar{\sigma}})T} \;,
\end{equation}
in which,
\begin{equation}
\begin{array}{cc}
I_{T}^\sigma=\int \varepsilon d\varepsilon 
\dfrac{N_\sigma(\varepsilon)\mathcal{G}_\sigma}{4k_BT\cosh^2(\dfrac{\varepsilon}{
2k_BT})} \;,
\\ \\
G_{\sigma}=\int d\varepsilon 
\dfrac{N_\sigma(\varepsilon)\mathcal{G}_\sigma}{4k_BT\cosh^2(\dfrac{\varepsilon}{
2k_BT})} \;.
\end{array}
\end{equation}
 
Now, we are able to calculate the charge and spin-dependent Seebeck coefficients in 
magnetic systems from the following equations:
\begin{equation}
\begin{array}{l}
S_{ch}=(S_\sigma+S_{\bar{\sigma}})/2 \;,
\\
S_{sp}=S_\sigma-S_{\bar{\sigma}} \;.
\end{array}
\end{equation}
To calculate these coefficients in our supposed systems we just need to 
determine the transmission probabilities via scattering processes.  

For applications in nano-scale cooling devices, it is useful to consider the 
figure of merit $\mathcal{Z}T$. In a magnetic system, the charge and spin 
currents can flow across the junction. So the charge and spin figures of merit 
are defined by
\begin{equation}
\begin{array}{cc}
\mathcal{Z}_{ch}T=\dfrac{G_{ch}S^2_{ch}T}{\kappa} \;, \\
\\
\mathcal{Z}_{sp}T=\dfrac{G_{sp}S^2_{sp}T}{\kappa_{sp}} \;, \\
\end{array}
\end{equation}
where $\kappa_{sp}=\mid\kappa_\sigma-\kappa_{\bar{\sigma}}\mid$ is the 
spin-polarized thermal conductivity whereas $G_{sp}= \mid 
G_\sigma-G_{\bar{\sigma}}\mid$ is the spin conductivity. The thermal 
conductivity in ballistic transport regime for each spin direction, is obtained 
after some straightforward algebra, by
\begin{equation}
\kappa_{\sigma}=\int \varepsilon^2d\varepsilon 
\dfrac{N_\sigma(\varepsilon)\mathcal{K}_\sigma}{4k_BT^2\cosh^2(\dfrac{\varepsilon}
{2k_BT})} \;.
\end{equation}

  \begin{figure}
   \includegraphics[width=8.0cm, height=5cm]{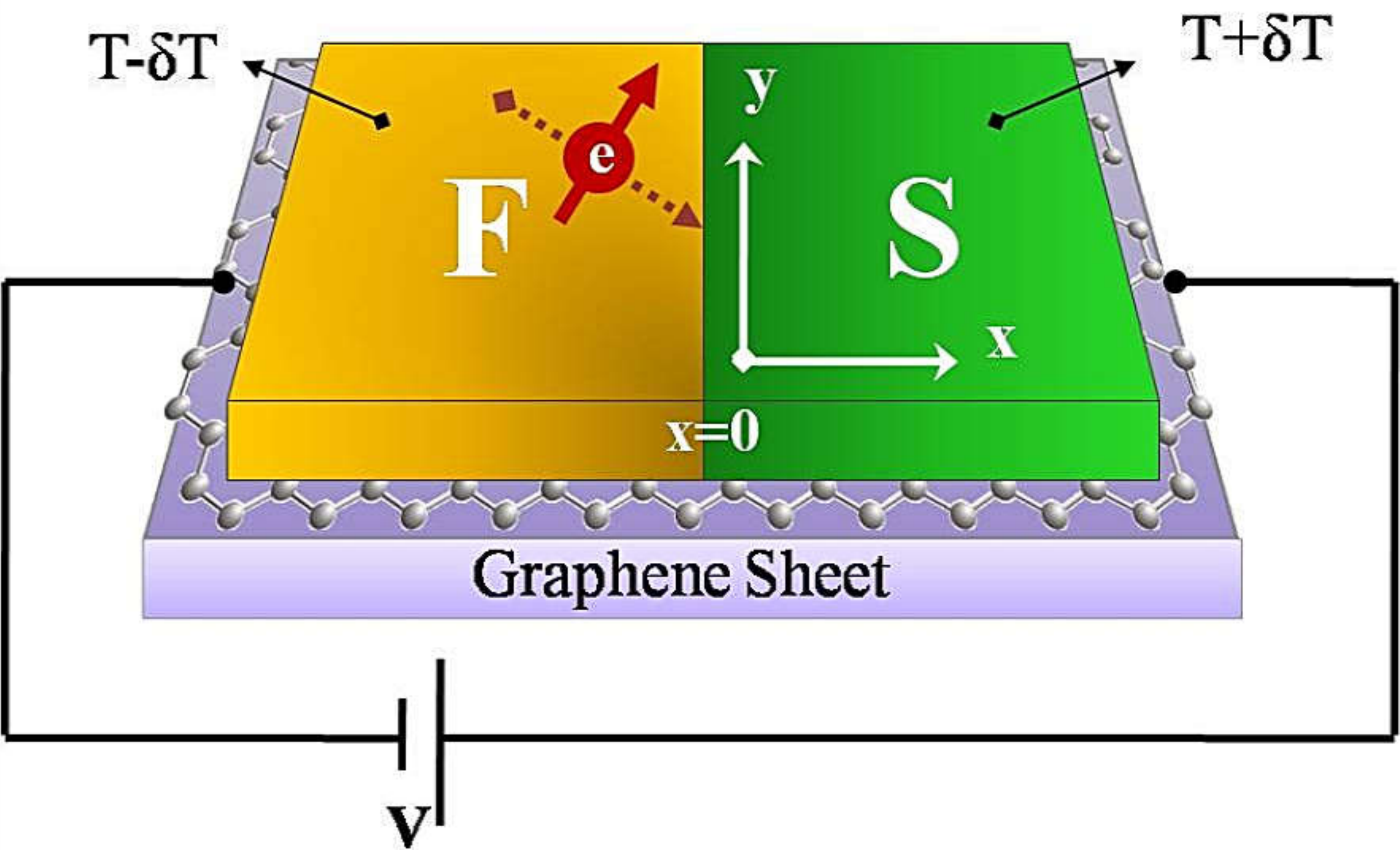}
   \caption{A schematic illustration of the graphene-based F-S junction. The 
      junction is set in the x-y plane and the uniform ferromagnetic and 
      superconducting regions are supposed semi-infinite. A possible path of an 
      incident spin-up electron at the F-S interface is also shown in the F 
      region. The F-S boundary is supposed at $x=0$.}
   \label{FS model}
  \end{figure}

\section{III. RESULTS AND DISCUSSIONS}
 \subsection{F-S Junction} 
We consider a F-S junction made of graphene as shown in Fig. \ref{FS model} 
where a gate voltage $V$ applied to the graphene sheet and keep their sides on 
different constant temperatures, $T-\delta T$ and $T+\delta T$. The superconducting and ferromagnetic features are induced in the graphene layer due to the proximity. To obtain the 
carrier's wave functions in graphene, we use the Dirac-Bogoliubov-de Gennes 
(DBdG) equation \cite{Beenakker2006,deGenne1966}
\begin{equation}
       \left(\begin{array}{cc}
       H-\mu^i & \Delta\\
       \Delta ^{\star} & \mu^i-{\mathcal{T}}[H]{\mathcal{T}}^{-1}\\
       \end{array}\right)
       \left(\begin{array}{c}
        u \\
        v \\
       \end{array}\right)=\varepsilon
       \left(\begin{array}{c}
       u \\
       v \\
       \end{array}\right)\;,
       \label{Eq.DBdG}
   \end{equation}
where $\mathcal{T}$ represents the time-reversal operator \cite{Beenakker2006}. 
$u $ and $v$ are the electron and hole parts of wave-functions, respectively. 
The term $\mu^i$ refers to the chemical potential of each region. The 
Hamiltonian of the F-S graphene junction 
$H=\mathcal{H}_\text{F}+\mathcal{H}_\text{S}+\mathcal{H}_\text{D}$, 
consist of ferromagnetic part $\mathcal{H}_\text{F} = 
\left(s_z\otimes\sigma_0\right) h$ for $x\leq 0$, superconductor part 
$\mathcal{H}_\text{S}=-U_0 s_0\otimes\sigma_0$ for $x\geq 0$ and the two dimensional 
Dirac Hamiltonian in one valley is $\mathcal{H}_D=s_0\otimes \hbar v_F \left(\sigma_x 
k_x+\sigma_y k_y\right)$ \cite{Beenakker2008}. In case of graphene, because of 
the valley degeneracy, the final results are multiplied by 2. In the above 
equations, $k_x$ and $k_y$ are the components of wave vector in $x$ and $y$ 
directions, respectively. $s_i$ and $\sigma_i$ are Pauli matrices, acting on 
real spin and pseudo-spin spaces of graphene. $s_0$ and $\sigma_0$ are $2\times 
2$ unit matrices, and for simplicity we assume $\hbar v_F=1$.

In the Hamiltonian of F segment, the magnetic exchange field $h$ is added 
to the Dirac Hamiltonian via the Stoner approach. For simplicity, we assume 
that $h$ is oriented in the $z$ direction without loss of generality 
\cite{Halterman2013}. This choice turns the exchange field to a good quantum 
number that allows for explicitly considering of $\uparrow$-spin and 
$\downarrow$-spin quasi-particles in F region and helps us to have more 
insightful analysis of spin-dependent phenomena in system. By diagonalizing the 
DBdG equation in the F region, eight eigenvalues are obtained as
  \begin{equation}
  \varepsilon=\pm \mu^\text{F} \pm\sqrt{(k^\text{F}_x)^2+(k_y)^2}\pm h \;.
  \end{equation} 
Therefore, the ferromagnetic exchange field alters the original energy value of 
electrons and holes according to their spin orientation.

  \begin{figure}
   \includegraphics[width=8.6cm, height=7cm]{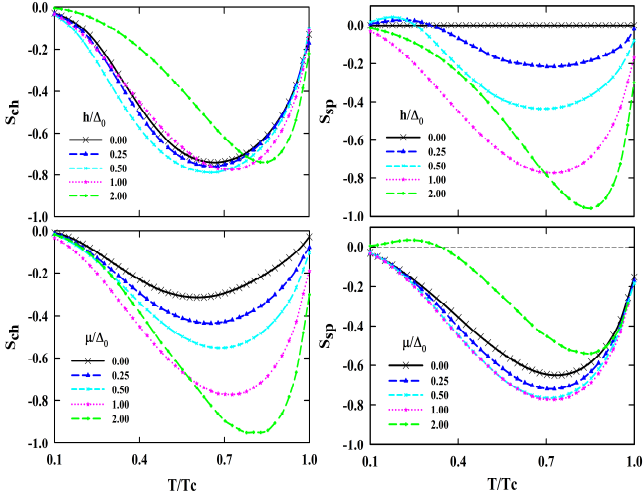}
   \caption{(Color online) The charge and spin thermopower of F-S junction as a 
    function of temperature. The chemical potential is assumed to be 
    $\mu=1.0\Delta_0$ in the top row and ferromagnetic exchange field values 
    are $h=\left({0.0,0.25,0.5,1.0,2.0}\right)\Delta_0$. In the 
    down panels, the applied magnetic exchange field is $h=1.0$ and the 
    chemical potential values are $\mu=\left({0.0,0.25,0.5,1.0,2.0}\right) 
    \Delta_0$. The temperature gradient is $T=0.1T_c$.}
   \label{vs T changing h and mu}
   \end{figure} 
The corresponding wave functions of the above eigenvalues are
\begin{equation}
 \begin{array}{l}
   \psi^{\text{F},\pm}_{e,\uparrow}(x)=\left(\mathbf{0^2}, 1 , \pm e^{\pm i
   \alpha_\uparrow^e}, \mathbf{0^4}\right)^{\bf T} e^{\pm i
   k_{x,\uparrow}^{\text{F},e} x} \;,
\\
   \psi^{\text{F},\pm}_{e,\downarrow}(x)=\left( 1 , \pm e^{\pm i 
   \alpha_\downarrow^e}, \mathbf{0^2},\mathbf{0^4}\right)^{\bf T} e^{\pm i 
   k_{x,\downarrow}^{\text{F},e} x} \;,
\\
   \psi^{\text{F},\pm}_{h,\uparrow}(x)= \left(\mathbf{0^4}, 1 ,  \mp
   e^{ \pm i \alpha_\uparrow^h},\mathbf{0^2}\right)^{\bf T} e^{ \pm
   i k_{x,\uparrow}^{\text{F},h} x} \;,
\\
   \psi^{\text{F},\pm}_{h,\downarrow}(x)=\left( \mathbf{0^4},
   \mathbf{0^2}, 1 , \mp e^{\pm i \alpha_\downarrow^h}\right)^{\bf T}
   e^{\pm i k_{x,\downarrow}^{\text{F},h} x} \;,
 \end{array}
\end{equation} 

where  $\mathbf{T}$ is a transpose operator and $\mathbf{0}^n$ denotes a 
$1\times n$ matrix with zero entries. From the conservation of the y-component 
wave vector ($k_y$) under the scattering processes, we factored out the 
corresponding multiplication \textit{i.e.}, $\exp(i k_y y)$. The 
$\alpha_{e,(h)}^{\uparrow(\downarrow)}$ variables are the propagation angles 
given by
   \begin{equation}
   \alpha_{\uparrow (\downarrow)}^{e(h)}=\arctan\left(\frac{ q_n}{k_{x, 
\uparrow(\downarrow)}^{\text{F},e(h)}} \right) \;.
   \end{equation}
We denote $k_y\equiv q_n$ that can vary in interval $-\infty \leq  q_n \leq 
+\infty$. The $e(h)$ superscript demonstrates the electron (hole)-like carriers
and $\uparrow (\downarrow)$ subscript refers to the spin direction. 

The superconductor region is assumed to be of s-wave symmetry type with 
step-like superconducting gap as,
\begin{equation}
    \Delta(x,T)=\begin{cases}
   0&       x\leq 0 \\
   \Delta(T)e^{i\phi} &  x\geq 0
    \end{cases},
    \label{Eq.7}
\end{equation}
where $\phi$ is the macroscopic phase of the superconductor. The 
temperature-dependent gap of superconductor, $\Delta(T)=\Delta_0 
\tanh(1.76\sqrt{(T_c/T)-1})$, deduced from the BCS theory \cite{Ketterson1999} 
in which $\Delta_0$ is the superconductor gap at zero temperature. In 
superconducting region, $U_0$ denotes the electrostatic potential that 
is very large ($U_0\gg 1$) in actual experiments compared to other system energy 
scales to satisfy the mean field requirements \cite{Beenakker2006}. We 
diagonalize the corresponding Hamiltonian and obtain the eigenvalues of S 
region as,
  \begin{equation}
  \varepsilon=\sqrt{|\Delta_0|^2+\Big(\mu^\text{S}+U_0\pm 
  \sqrt{(k_x^\text{S})^2+(k_y)^2}\Big)^2} \;.
  \label{Eq.17}
  \end{equation}
The associated wave functions are
   \begin{equation}
   \begin{array}{l}
   \psi^{\text{S},\pm}_{e, \kappa=1}(x)=\Big(e^{+i \beta}, \pm e^{+i\beta}, 
    \mathbf{0^2}, e^{-i\phi}, \pm e^{-i\phi},\mathbf{0^2}\Big)^{\bf T}e^{\pm i 
    k_x^{\text{S},e} x} \;,
\\
   \psi^{\text{S},\pm}_{e,\kappa=2}(x)=\Big (\mathbf{0^2}, e^{+i \beta}, \pm 
    e^{+i \beta}, \mathbf{0^2}, e^{-i\phi}, \pm e^{-i\phi}\Big)^{\bf T}e^{\pm i 
    k_x^{\text{S},e} x} \;,
\\
   \psi^{\text{S},\pm}_{h, \kappa=1}(x)= \Big (e^{-i \beta}, \mp e^{-i \beta}, 
   \mathbf{0^2}, e^{-i\phi}, \mp e^{-i\phi}, \mathbf{0^2}\Big)^{\bf T}e^{\pm i 
   k_x^{\text{S},h} x} \;,
\\
   \psi^{\text{S}, \pm}_{h, \kappa=2}(x)= \Big (\mathbf{0^2}, e^{-i \beta}, \mp 
    e^{-i \beta}, \mathbf{0^2}, e^{-i\phi}, \mp e^{-i\phi}\Big)^{\bf T}e^{\pm i 
    k_x^{\text{S},h} x} \;.
   \end{array}
   \label{WF.S}
   \end{equation} 
The parameter $\beta$ is responsible for electron-hole conversions at F-S 
interface that depends on the temperature-dependent superconducting gap: 
    \begin{equation}
    \beta=\begin{cases}
    +~\arccos(\varepsilon/\Delta)&       \varepsilon\leq \Delta \\
    -i ~\text{arccosh}(\varepsilon/\Delta) &   \varepsilon \geq\Delta
    \end{cases}
    \label{Eq.19}.
    \end{equation}


 \begin{figure}
 \includegraphics[width=8.6cm, height=6.8cm]{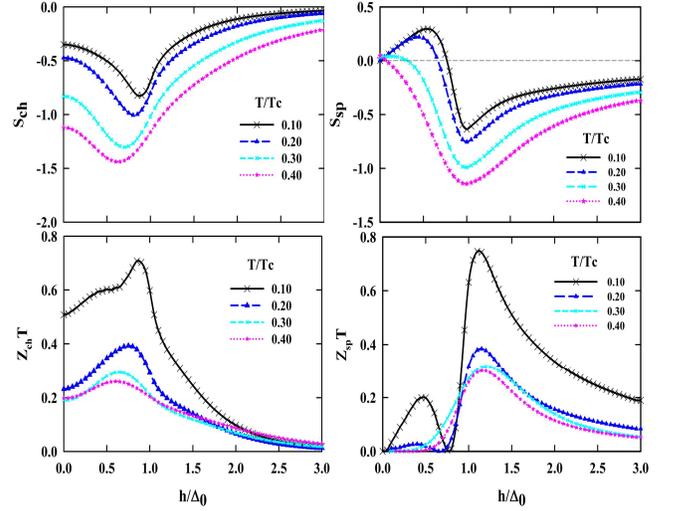}
  \caption{(Color online) The charge and spin thermopower and figures of merit 
  of F-S junction as a function of the magnetic exchange field. The applied 
  chemical potential is considered $\mu=1.0\Delta_0$ and the temperature 
  gradient values are $T=\left(0.1,0.2,0.3,0.4 \right)T_c$.}
  \label{vs h changing T}
 \end{figure}

We normalize energies by the superconductor gap at zero temperature $\Delta_0$ 
and lengths by the superconducting coherent length $\xi_S=\hbar v_F/\Delta_0$.

In the F region, we assume that a right moving electron with $\uparrow$-spin 
direction hits the interface of F-S. We match the obtained wave-functions at the 
boundary $x=0$, and calculate the charge and thermal transmissions through 
normal and Andreev reflections at the interface from following equations:
\begin{equation}
 \begin{array}{cc}
  r_{N\uparrow\downarrow}=\dfrac{-\cos\beta\sin(\alpha^h_{\downarrow}+\alpha^e_{
  \uparrow})/2+i\sin\beta\cos(\alpha^h_{\downarrow}-\alpha^e_{ 
  \uparrow})/2}{\cos\beta\cos(\alpha^h_{\downarrow}-\alpha^e_{\uparrow} 
  )/2+i\sin\beta\cos(\alpha^h_{\downarrow}+\alpha^e_{\uparrow})/2}\;,
  \\\\
  r_{A\uparrow\downarrow}=\dfrac{e^{-i\phi}\sqrt{\cos\alpha^{e}
  _\uparrow\cos\alpha^{h}_\downarrow}}{\cos\beta\cos(\alpha^h_{\downarrow}
  -\alpha^e_{\uparrow} )/2+i\sin\beta\cos(\alpha^h_{\downarrow}+ 
  \alpha^e_{\uparrow})/2}\;.
 \end{array}
\end{equation}
The thermoelectric properties of the junction are obtained using these 
equations. In the top row of Fig. \ref{vs T changing h and mu}, we demonstrate 
the temperature dependent charge and spin-dependent Seebeck coefficients ($S_{ch}, 
S_{s}$) for different values of normalized magnetic exchange field 
($h/\Delta_0$). 

 \begin{figure}
  \includegraphics[width=9cm, height=5cm]{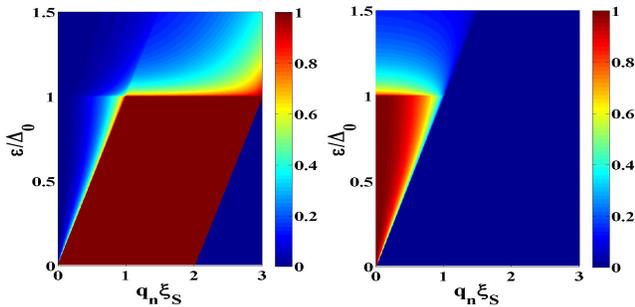}
  \caption{(Color online) The probability of normal 
    $|r_{N}^\uparrow|^2$ (left) and Andreev $|r_{A}^\downarrow|^2$(right) 
    reflections of F-S in low-doping regime with $\mu=h=1.0\Delta_0$ in which 
    the conventional normal reflection is dominate. The temperature is 
    $T=0.1T_c$.}
   \label{FS probabilities}
 \end{figure}
   
We know from the previous calculations \cite{Rameshti2015,Ghosh2015} that in the magnetic 
graphene, the strong thermoelectric effects happened in the intermediate 
temperatures when the $k_BT$ parameter is in the order of $\mu+h$. In the 
superconductor graphene-based junction, we have two separate regimes: i) Weak 
field regime with $h\leq\Delta_0$ in which charge and spin-dependent Seebeck coefficients 
have a non-linear relation with temperature gradient and increase with 
increasing $h$. This fact is the consequence of the existence of ferromagnetic 
exchange potential. It is not only leads to an imbalance between the population 
of $\uparrow$- and $\downarrow$- spins in graphene sheet, but also can affect 
the amplitudes of Andreev reflections. However, in this regime the graphene is 
doped ($\mu\neq 0$)and both spin and charge currents exist while the second 
dominates. For $h\leq\mu$, increasing Andreev reflections leads to 
a non-linear increase in thermoelectric effects. ii) $h\geq\Delta_0$ in which 
increasing the chemical potential at higher temperatures leads to decreasing the 
spin thermopower and enhancing the charge thermopower. Since in graphene-based 
systems the doping can be controlled easily, one can reach the regimes in which 
the Fermi energy is comparable to thermal energy $k_BT$ to have a more efficient 
values of Seebeck coefficients (See the bottom row of Fig. \ref{vs T changing h 
and mu}). Moreover, at very low temperatures, $T/T_c\ll \Delta_0$, the values of 
charge and spin-dependent Seebeck coefficients are very small and go to zero with respect 
to the temperature.
  \begin{figure}
   \includegraphics[width=8.6cm, height=6.8cm]{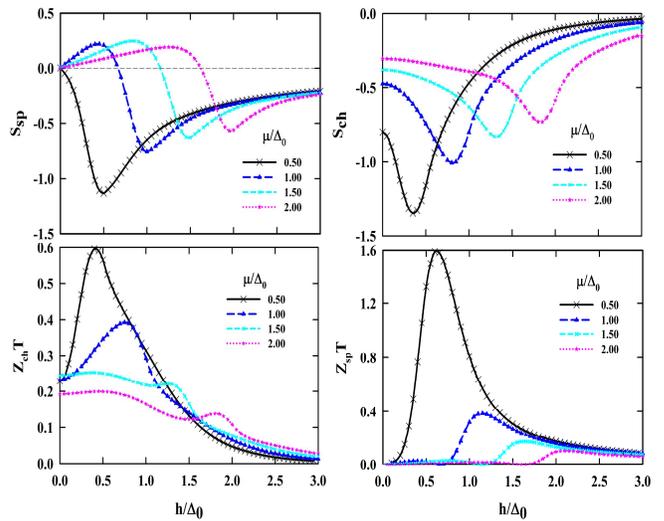}
   \caption{(Color online) The charge and spin-dependent Seebeck and figures of merit of 
    F-S junction as a function of normalized magnetic exchange field. The 
    chemical potential values are $\mu=\left(0.5,1.0,1.5,2.0 
    \right)\Delta_0$ and the temperature is equal to $0.1 T_c$.}
  \label{vs h changing mu}
  \end{figure}
 
In our subsequence analysis we study the dependence of the thermoelectric 
coefficients on the value of magnetic exchange field in Fig. \ref{vs h changing 
T}. For some small values of $h$ and low enough temperature, the spin-dependent Seebeck 
effect has positive value while the charge Seebeck effect becomes all negative. 
We know from the dispersion relation of ferromagnetic graphene that in the 
low-doping regime ($\mu\leq\Delta_0$), the $\uparrow$-spin electrons and 
$\downarrow$-spin holes in the conduction and valance bands of graphene 
accompany each other to build the spin-dependent Seebeck effect. At low magnetic 
exchange field, the temperature gradient caused carriers with different spin 
directions to compete with each other resulting in an spin accumulation.
For better analysis, the probabilities of normal ($|r_{N}^\uparrow|^2$) and 
Andreev $(|r_{A}^\downarrow|^2)$ reflections are shown in Fig. \ref{FS 
probabilities}. It can be easily conclude that the $\uparrow$-spin carriers 
dominate the spin-dependent Seebeck coefficient and since they carry negative current, the 
$S_s$ is negative, regardless to the strength of spin-splitting (Fig. \ref{FS 
probabilities}). In low-temperature regime ($T/T_c\leq 0.2$) there is a sign 
change in the spin thermopower. But at higher temperatures, the spin-dependent Seebeck 
coefficient becomes all negative dominated by majority $\uparrow$-spin carriers.

Experimentally, by tunning the chemical potential, the charge carriers of 
graphene can be easily controlled. Consequently, the Fermi level in graphene can 
be electron-like or hole-like. Moreover, the magnetic exchange field in 
ferromagnetic region also changed the Fermi level of carriers. We use this fact 
for controlling the sign of spin-dependent thermopower of the junction, as shown 
in Fig. \ref{vs h changing mu}. In low-doped regime, the amplitude of the 
charge (spin) Seebeck coefficients and figures of merit is large because the 
thermal transport is very weak but the thermoelectric effect is not. These 
coefficients also vary by changing the gate voltages due to the changing in the 
transmission probabilities. As it is expected, the maximum values of spin 
(charge) figure of merit occurs at some $h$ above (below) the value of $\mu$.


 \begin{figure}
    \includegraphics[width=8cm, height=5cm]{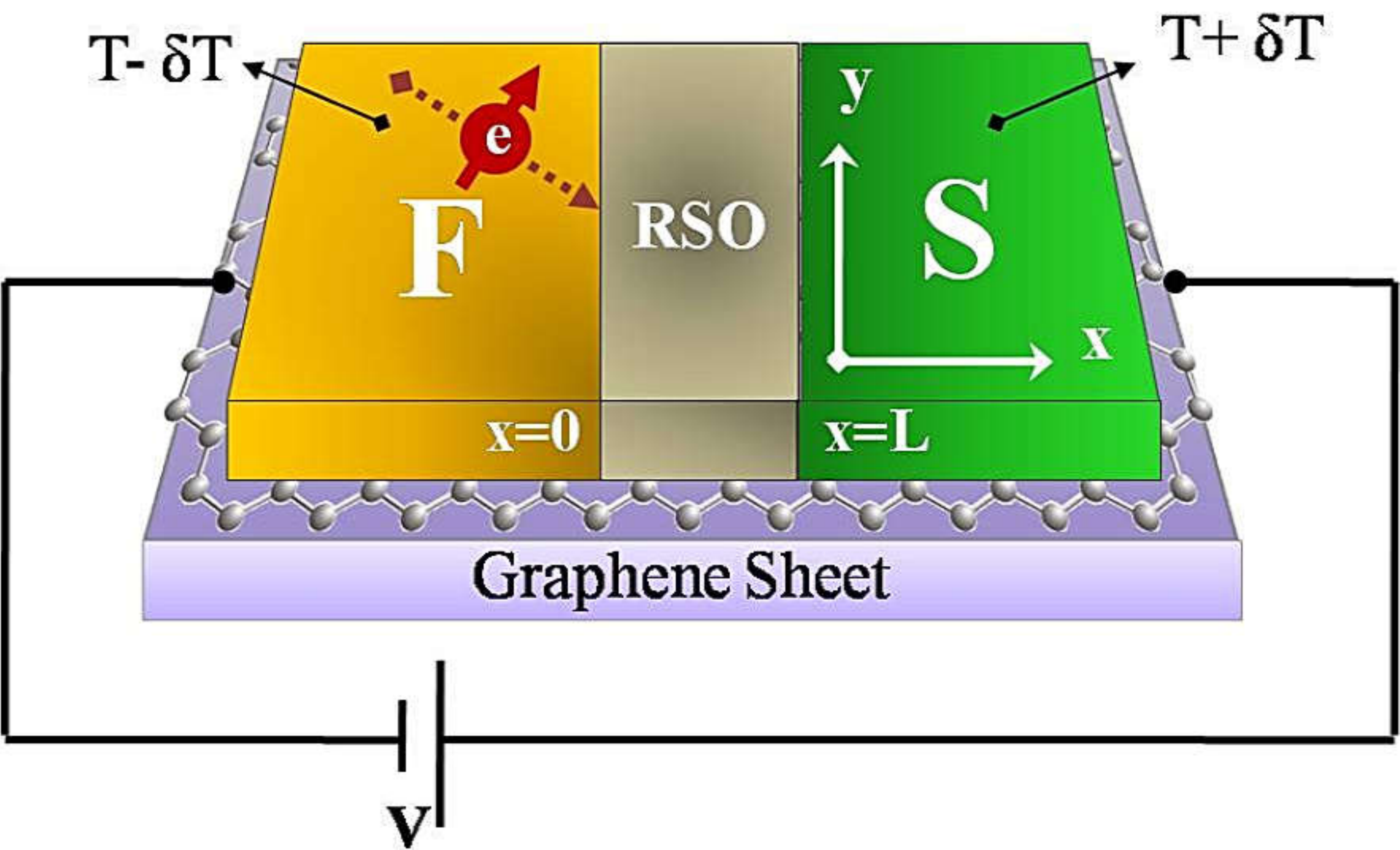}
    \caption{Schematic illustration of a graphene-based F-RSO-S junction. 
    Similar to Fig.\ref{FS model}, the junction is set in the x-y plane and a 
    possible path of an incident spin-up electron at the F-RSO interface is 
    also shown in the F region. The boundaries are considered at $x=0$ and 
    $x=L$.}
    \label{FRSOS model}
 \end{figure}

\subsection{F-RSO-S Junction}
It is known that the imposition of the thermal gradient implies an equilibrium 
between electrons and their counterparts. This effect along with ferromagnetic 
features produce spin thermopower in the junction, as discussed in the previous 
section. Now, we set the spin-mixer barrier with length of L in the middle of 
ferromagnetic and superconductor regions of graphene as shown in Fig. 
\ref{FRSOS model}. The importance of this suggestion is that the spin-mixing 
phases in the proximity-induced RSO region provides singlet to triplet conversion 
\cite{Beiranvand2016} in the heterostructure gives rise to an spin transport 
mechanism through ferromagnetic region with strong exchange splitting of its 
bands. In this case, the Hamiltonian $H$, in Eq. \ref{Eq.DBdG} consists of two 
parts named $\mathcal{H}_D$ and $\mathcal{H}_\text{RSO}=-\lambda \left( 
s_y\otimes\sigma_x-s_x\otimes\sigma_y\right)$ in which $\lambda$ is the Rashba 
spin-orbit parameter. By diagonalizing the Hamiltonian of RSO region, we find 
the following dispersion relation:
 \begin{equation}
 \begin{array}{cc}
  \varepsilon=\pm\mu^\text{RSO}+\zeta \sqrt{(k_x^\text{RSO})^2+(k_y)^2+
  \lambda^2}+ \eta \lambda \;, 
  \label{EigenHSO}
 \end{array}
 \end{equation}
where $\eta,\zeta=\pm 1 $ indicate band indices. In contrast to intrinsic spin 
orbit couplings, the energy spectrum in the presence of RSO is gap-less with a 
splitting of magnitude $2\lambda$ between sub-bands. This sub-band splitting 
results in interesting phenomena \cite{Beiranvand2016}. The wave-functions 
associated with the eigenvalues can be expressed by: 
 \begin{flushleft}
  \begin{equation}
  \begin{array}{l}
  \psi^{\text{RSO},\pm}_{e,\eta=+1}(x)= \Big (\mp i f_{+}^e e^{\mp 
    i\theta_{+}^e}, -i, 1, \pm f_{+}^e e^{\pm i
    \theta_{+}^e},\mathbf{0^4}\Big)^{\bf T}e^{\pm i k_{x+}^{e}x} \;,
\\
  \psi^{\text{RSO},\pm}_{e, \eta=-1}(x=\Big(\pm f_{-}^e e^{\mp i \theta_{-}^e}, 
1, -i, \mp if_{-}^e e^{\pm i \theta_{-}^e},\mathbf{0^4}\Big)^{\bf T}e^{\pm i 
k_{x-}^{e}x} \;,
\\
  \psi^{\text{RSO},\pm}_{h,\eta=+1}(x)=\Big (\mathbf{0^4},\mp i f_{+}^h e^{\mp 
i \theta_{+}^h}, -i, 1, \pm f_{+}^h e^{\pm i \theta_{+}^h}\Big)^{\bf T}e^{\pm i 
k_{x+}^{h}x} \;,
\\
  \psi^{\text{RSO}, \pm}_{h, \eta=-1}(x)=\Big (\mathbf{0^4}, \pm f_{-}^h e^{\mp 
    i \theta_{-}^h}, 1, -i, \mp if_{-}^h e^{\pm i \theta_{-}^h}\Big)^{\bf 
    T}e^{\pm   i   k_{x-}^{h}x} \;.
  \end{array}
  \end{equation}
 \end{flushleft}

  \begin{figure}
    \includegraphics[width=10.6cm, height=6.8cm]{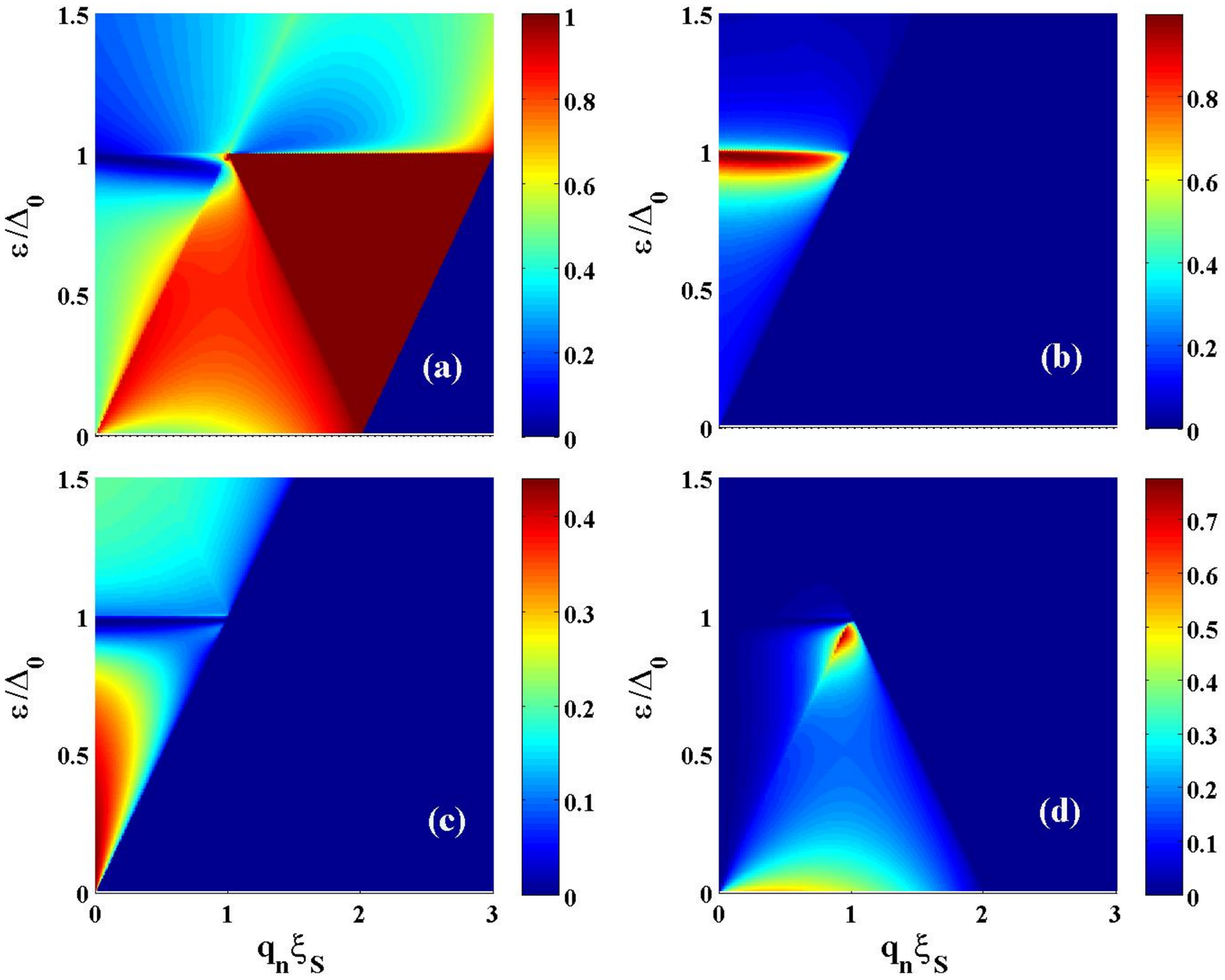}
    \caption{(Color online) Back scattering probabilities of an incident 
      $\uparrow$-spin particle at $x=0$ as a function of applied voltage 
      $\varepsilon/\Delta_0$ across the junction and the transverse momentum 
      $q_n\xi_S$ for (a) conventional normal reflection $|r_{N}^\uparrow|^2$, 
      (b) conventional Andreev reflection $|r_{A}^\downarrow|^2$, (c) 
      spin-flipped normal reflection $|r_{N}^\downarrow|^2$, and (d) 
      anomalous Andreev reflection $|r_{A}^\uparrow|^2$. The input values are 
      $\mu=h=1.0\Delta_0$, $T=0.1T_c$ and $\lambda=5.0 \Delta_0$.}
    \label{FRSOS Probabilities}
  \end{figure}  
     
Similar to the F and S region, each wave functions in the RSO region must be 
multiplied to $e^{i q_ny}$, where omitted here because of simplicity.

The definition of auxiliary parameters are:   
\begin{equation}
    \begin{array}{cc}
      f_\eta^e=\sqrt{1+2\eta\lambda(\mu^\text{RSO}+\varepsilon)^{-1}} \;, \\
      f_\eta^h=\sqrt{1+2\eta\lambda(\mu^\text{RSO}-\varepsilon)^{-1}} \;,
    \end{array}
\end{equation} 
and  
\begin{equation}
  \begin{array}{cc}
  \theta_{\eta}^e=\arctan\left(q_n/k_{x,\eta}^e\right) \;, \\
  \theta_{\eta}^h=\arctan\left(q_n/k_{x,\eta}^h\right) \;,
    \end{array}
\end{equation}
where $\theta_{\eta}^{e}$ and $\theta_{\eta}^{h}$ are the electron and 
hole propagation angles in the region with spin-orbit interaction. We note that 
if the transverse component of wave-vector goes beyond a critical value $q^c$, 
the wave-functions turn to evanescent modes. Here, however, since the RSO region 
is confined between F and S regions, the evanescent modes contribute to the 
quantum transport process. We thus take both the propagating and decaying modes 
into account throughout our calculations. By matching the wavefunctions at the 
interfaces, \textit{i.e.}, $\Psi^\text{F}(x)=\Psi^\text{RSO}(x)$ at $x=0$ and 
$\Psi^\text{RSO} (x)=\Psi^\text{S} (x)$ at $x=L$, we obtain all of the 
scattering coefficients \cite{Beiranvand2016}.

 \begin{figure}
 \includegraphics[width=8.6cm,height=6.8cm]{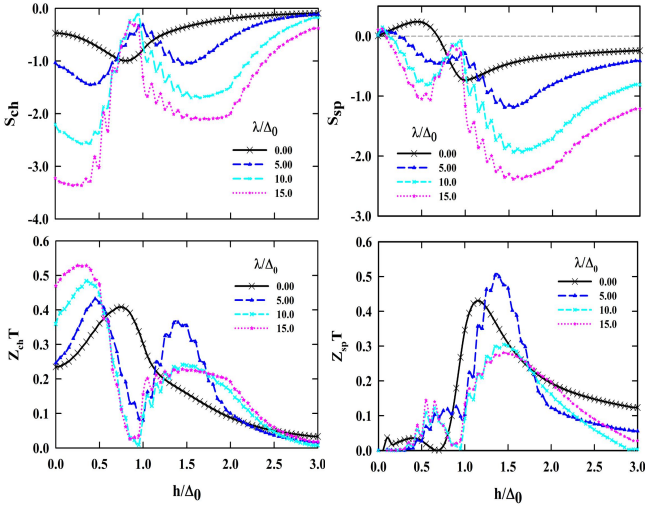}
   \caption{(Color online) The charge and spin-dependent Seebeck coefficients and figures 
    of merit of F-RSO-S junction as a function of normalized $h$. $T=0.1 T_c$ 
    and $\mu=1.0\Delta_0$.}
  \label{vs h changing lambda}
 \end{figure}
 
In this case, the transmission probabilities have more complicated 
energy-dependent form. So, the results could not report analytically. Consider, 
for instance, the four scattering processes of an incident $\uparrow$-spin 
electron arriving at the RSO interface from the ferromagnetic side. For 
$\lambda=0$, the intermediate region acts like a normal graphene. But, at 
nonzero values of $\lambda$, it can undergo an spin-mixing process in the RSO 
segment. For $\varepsilon\leq\Delta$, the incident particle cannot penetrate 
deep into the superconductor and gets reflected back into the ferromagnetic 
region either in the form of an electron (specular reflection) or, 
alternatively, as a hole (Andreev reflection). In this case, in addition to the 
normal reflection $(r_{N}^\uparrow)$ and the conventional Andreev reflection 
$(r_A^\downarrow)$, the probabilities of anomalous Andreev reflection 
($r_{A}^\uparrow$) and unconventional normal reflection ($r_{N}^\downarrow$) 
have finite amplitudes \cite{Beiranvand2016}. If $\lambda$ is chosen properly, 
one can be able to reach an approximately high value of thermopower coefficients 
through applying spin-orbit interaction in the system. 

The numerically obtained results for conventional and anomalous normal and 
Andreev reflections are presented in Fig. \ref{FRSOS Probabilities} in four 
different panels. It should be noted that, in order to achieve a high value of 
spin-dependent Seebeck coefficient with the possibility of using in the 
spin-thermoelectric devices, the length of suggested system should be in the 
order of 1 $\mu m$ \cite{Guimaraes2014}. Since the device length is normalized 
to the superconducting coherence length ($\xi_S$), all of our supposed 
structures are easily applicable in current experimental setups. Furthermore, in 
graphene, the Rashba spin-orbit coupling is in the order of $\approx$ 17 meV and 
the typical superconducting gap is about $0.1-3$ meV. So, in the theoretical 
point of view, we can easily increasing the RSO parameter ($\lambda$) up to 15 
times greater than the value of $\Delta_0$ and the results can be experimentally 
reliable.

Since the Rashba spin-orbit interaction provides a spin mixing procedure in the 
middle region, one can find appropriate amounts of $\mu$, $h$ and $\lambda$ in 
which the charge and spin currents reach to their critical values. In Fig. 
\ref{vs h changing lambda}, the numerically obtained results for charge 
and spin-dependent Seebeck coefficients are presented as a function of $h$. In this 
figure, we show the possibility of enhancing the spin-dependent thermopower in 
graphene devices by spin-orbit engineering to a large value of 2.5$k_Be$. The 
maximum value of the spin thermopower ($S^{max}_s$) was shown to increase with 
increasing the Rashba parameter. Remarkably, with a magnetic exchange field of 
$h\approx 0.1\Delta_0$, the charge Seebeck coefficient can also reach a value 
greater than 3.0$k_Be$, which is approximately 15 times larger than the value 
in pristine graphene.
\begin{figure}
    \includegraphics[width=8.6cm,height=10cm]{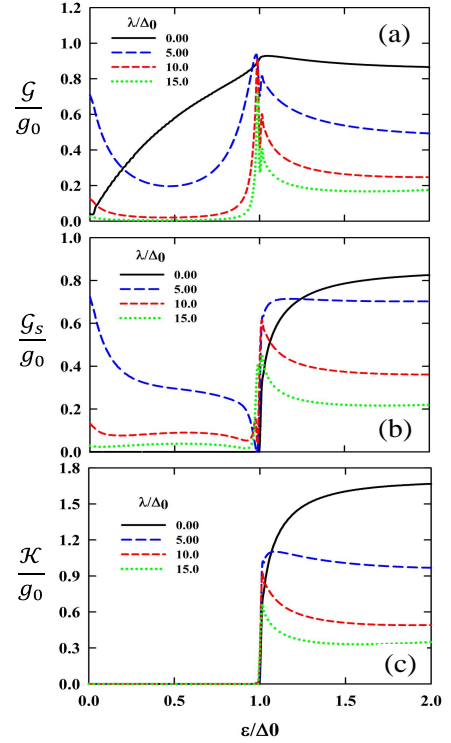}
    \caption{(Color online) The normalized (a) charge, (b) spin and (c) thermal 
    conductances versus normalized energy. $T=0.1 T_c$ and $\mu=h=1.0 
    \Delta_0$.}
    \label{conductance}
\end{figure}
Furthermore, we observe a critical zero point in which both charge and spin 
Seebeck effects drop down to zero. The origin of this effect in the low-doped 
regime can be understood from transmission coefficients. In this regime 
($\mu\approx\Delta_0$), when the RSO parameter is zero ($\lambda=0$), we have 
normal region with finite width instead of RSO region which is the origin of 
resonant tunneling. In this case, there are only two reflection probabilities 
for a $\uparrow$-spin electron hitting the interface: ($a$) conventional normal 
reflection ($r_{N}^\uparrow$) and ($b$) retro Andreev reflection ($\varepsilon 
\leq \Delta_0$) or specular Andreev reflection ($\varepsilon \geq \Delta_0$), 
depending on the quasi-particle's energy. Tuning the system parameters, 
($\mu\approx \Delta_0$), $ \lambda \neq 0$ lead to spin-mixing process in the 
RSO segment. This effect is responsible to tiny oscillations in the charge and 
spin-dependent Seebeck curves (See Fig. \ref{vs h changing lambda}). This spin 
mixing process will increase the amplitudes of resonant oscillations. 
In addition, the transmission probability spectrum drops to zero for energy 
corresponding to the Fermi level position which are known as specular AR limit. 
In this case, the conventional Andreev reflected hole is placed in the valence 
with a specular reflection process whereas the anomalous one passes through 
the conduction band with a retro-reflected process. Thus, the reflected
holes separate from each other with respect to their spin directions  
\cite{Beiranvand2016}. This suggests that there is a potential for application 
of this setup for cooling in the spin-dependent apparatuses.

Another experimentally measurable quantity in our supposed systems is the 
conductance of the junction. The behavior of charge, spin and thermal 
conductances are shown in Fig. \ref{conductance}. The presence of exchange 
energy $h$ in the F region results in an imbalance between carriers. In the 
presence of RSO coupling $\lambda \neq 0$, however, due to the possibility of 
spin-mixing in this region, the spin-polarized conductance can be also nonzero 
in the sub-gap region $\varepsilon\leq\Delta_0$. The enhancement of sub-gap spin 
conductance also traces back to the generation of equal-spin triplet 
correlations as discussed previously \cite{Beiranvand2016}.
It is obvious from Eq. (\ref{EigenHSO}) that further increase in the value of RSO parameter leads to large band splitting between the $\eta=1$ and $\eta=-1$ bands in the RSO region. Owing to this effect contribution of $\eta=1$ band in transport mechanism is negligible and leads to reduction of the values of the spin, charge and thermal conductances of the junction.

\section{Conclusion}
To conclude, we have developed a theoretical framework to study the thermal 
transport properties of ferromagnetic-superconductor (F-S) and 
ferromagnetic-Rashba spin-orbit-superconductor (F-RSO-S) junctions by 
calculating charge and spin-dependent thermopower of the junctions. Our results 
demonstrate that the simultaneous effect of spin-splitting field and spin orbit 
coupling leads to a strong suppression of the spin imbalances which can 
affect on the usually small thermoelectric coefficients in superconducting 
nanostructures. The results show that in the low-doped regime, by engineering 
the spin-orbit interaction, the charge and spin thermopowers can reach to 
their maximums up to 3.0$k_B/e$ and 2.5$k_B/e$, respectively in the F-RSO-S 
junction. Remarkably, these coefficients have a zero-point critical value with 
respect to magnetic exchange field and chemical potential. This effect 
disappears when the Rashba coupling is absent. These results suggest that 
graphene-based superconductors can be used in spin-caloritronic devices.

\section{Acknowledgment}
The authors acknowledge the Iran national science foundation (INSF) for financial support.

\end{document}